\documentclass[cmp]{svjour}
\usepackage{amssymb}
\usepackage{amsfonts}
\newcommand{\trace}{\mathop{\rm Tr}\nolimits}

\newcommand{\ran}{\mathop{\rm ran}\nolimits}

\newcommand{\bra}[1]{\langle#1|}
\newcommand{\ket}[1]{|#1\rangle}

\newcommand{\ktimes}{\boxtimes}
\newcommand{\cB}{{\cal B}}

\newcommand{\cH}{{\cal H}}

\newcommand{\cS}{{\cal S}}

\newcommand{\R}{{\mathbb{R}}}
\newcommand{\identity}{{\mathbb{I}}}

\newcommand{\be}{\begin{equation}}
\newcommand{\ee}{\end{equation}}
\newcommand{\bea}{\begin{eqnarray}}
\newcommand{\eea}{\end{eqnarray}}
\newcommand{\beas}{\begin{eqnarray*}}
\newcommand{\eeas}{\end{eqnarray*}}
\begin{document}
\title{On Strong Subadditivity of the Entanglement of Formation}
\author{Koenraad M.R. Audenaert \and Samuel L. Braunstein}
\institute{University of Wales, Bangor \\
School of Informatics \\
Bangor (Gwynedd) LL57 1UT, Wales \\
\email{kauden@informatics.bangor.ac.uk}
}
\date{\today}
\def\makeheadbox{}
\maketitle
\begin{abstract}
We employ a basic formalism from convex analysis to show a simple relation between
the entanglement of formation $E_F$ and the conjugate function $E^*$ of
the entanglement function $E(\rho)=S(\trace_A\rho)$.
We then consider the conjectured strong superadditivity of the entanglement of formation
$E_F(\rho) \ge E_F(\rho_I)+E_F(\rho_{II})$, where $\rho_I$ and $\rho_{II}$ are the
reductions of $\rho$ to the different Hilbert space copies, and prove that it is equivalent with subadditivity
of $E^*$. Furthermore, we show that strong superadditivity would follow from
multiplicativity of the maximal channel output purity for
quantum filtering operations, when purity is measured by Schatten $p$-norms for $p$ tending to 1.
\end{abstract}
\section{Introduction} \label{intro}
One of the central quantities in quantum information theory is the entanglement cost of a state, defined as
the number of maximally entangled pairs (singlets) required to prepare this state in an asymptotic way.
Calculating the entanglement cost of a general mixed state as such is, with the present state of knowledge,
a formidable task because one has to consider an infinite supply of singlets and construct a protocol using
local or classical (LOCC) operations only, such that the resulting (infinite-dimensional) state approximates an
infinite supply of the required state to arbitrary precision. Furthermore, the protocol must have maximal yield,
the number of states produced per singlet. The entanglement cost is the inverse of this yield.

An important theoretical breakthrough was achieved in \cite{hayden}, where the entanglement cost $E_C$
was shown to be equal to the regularised entanglement of formation:
$E_C(\rho) = \lim_{n\rightarrow\infty} E_F(\rho^{\otimes n})/n$. The entanglement of formation (EoF)
(defined below in (\ref{eq:def_eof})) is defined in a mathematical and non-operational way and is
therefore much more amenable to calculation.
Moreover, for 2-qubit mixed states, a closed formula for the EoF exists \cite{wootters}.
Nevertheless, calculating the entanglement cost still requires calculations over infinite-dimensional states.
For that reason one would hope for the additivity property to hold for the EoF:
$E_F(\rho_1\otimes\rho_2) =?\, E_F(\rho_1)+E_F(\rho_2)$, because then $E_C=E_F$.
Additivity of the EoF has been proven in specific instances \cite{benatti,kalle,vidal,sen,winter02,fan}.
Some of these additivity results are sufficiently powerful to allow calculating the entanglement cost
for certain classes of mixed states \cite{vidal,sen,winter02}.
The much sought-after general proof, however, remains elusive for the time being and, in fact, general
additivity is still a conjecture.

It is very easy to show that the EoF is \textit{subadditive}:
$$
E_F(\rho_1\otimes\rho_2) \le E_F(\rho_1)+E_F(\rho_2).
$$
Additivity would then follow from \textit{superadditivity}:
$$
E_F(\rho_1\otimes\rho_2) \ge?\, E_F(\rho_1)+E_F(\rho_2).
$$
In \cite{kalle} a stronger property, which would imply (super)additivity, has been conjectured
for the EoF, namely \textit{strong superadditivity}:
$$
E_F(\rho) \ge?\, E_F(\rho_I)+E_F(\rho_{II}),
$$
where $\rho$ is a general state over a duplicated Hilbert space and $\rho_I$ and $\rho_{II}$
are its reductions to the different copies of that space.

In this paper we show that strong superadditivity of EoF is equivalent to subadditivity of a much simpler
quantity, the so-called \textit{conjugate} of the entanglement functional $E(\rho)=S(\trace_A\rho)$.
We then exploit this equivalence to show that strong superadditivity
would follow as a consequence of multiplicativity of the
\textit{maximal output purity}, measured by a Schatten norm, for quantum filtering operations
(this quantity will also be defined in due course).

The main results are stated in Theorems 1 and 2.
To arrive at these results, we have made use of a basic formalism from convex analysis
\cite{rock,boyd} and we hope that our results will stimulate usage of this
elegant theory in other areas of quantum information.
%
\section{Notations} \label{sec:not}
Let us first introduce the basic notations.
Let $S(\rho)$ denote the von Neumann entropy $S(\rho) = -\trace\rho\ln\rho$.
For state vectors we will typically use lowercase Greek letters, $\psi$, $\phi$, and assign the corresponding
uppercase letter to the projector of that vector; e.g.\ $\Psi=\ket{\psi}\bra{\psi}$.
For mixed states we will use lowercase Greek letters $\rho$, $\sigma$, $\tau$.
The identity matrix will be denoted by $\identity$.

We shall denote the set of bounded Hermitian operators over the Hilbert space $\cH$ by $\cB^s(\cH)$,
the set of non-negative elements in $\cB^s(\cH)$ by $\cB^+(\cH)$,
and the (convex) set of all states (trace 1 positive operators) over $\cH$
by $\cS(\cH)$.

We will frequently slim down expressions
like $\max_{\rho\in\cS} \{\ldots\}$ to $\max_\rho \{\ldots\}$. When the domain of, say, a maximisation
over states is missing it will be implicitly understood that the whole of state space $\cS(\cH)$ is meant.
The abovementioned
naming convention for states and vectors will be adhered to exactly for that reason.

Any state $\rho$ can be realised by an ensemble of pure states.
An ensemble is specified by a set
of pairs $\{(p_i, \psi_i)\}_{i=1}^N$, consisting of $N$ state vectors $\psi_i$ and associated statistical weights $p_i$
(with $p_i\ge 0$ and $\sum_i p_i=1$). Here, $N$ is called the cardinality of the ensemble.
The entanglement of formation (EoF) of a bipartite state $\rho$
(i.e., a state over the bi-partite Hilbert space $\cH_A\otimes \cH_B$),
is defined by \cite{bennett}
\be \label{eq:def_eof}
E_F(\rho) = \min_{\{(p_i,\psi_i)\}} \big\{ \sum_i p_i S(\trace_B \Psi_i) : \sum_i p_i \Psi_i = \rho \big\}.
\ee
\section{Convex Closures} \label{sec:conv}
Admittedly, the definition of the EoF just mentioned is not very handy to work
with. Not in the least because for generic states $\rho$ the cardinality $N$
of the optimal realising ensemble must be larger than $R^{1.5}/4$, where $R$ is the rank of $\rho$ \cite{lockhart}.
This is one of the reasons why no really efficient numerical algorithms have been found yet to calculate
the EoF \cite{ka}. Furthermore, the mere fact that the minimisation involves ensembles at all makes a theoretical
study of the EoF rather difficult. One of the first attempts at proving additivity of EoF relied
on the investigation of these optimal ensembles \cite{benatti}.

The results in the present work depend on the following simple observation.
The import of the definition (\ref{eq:def_eof}) of the EoF, as has been shown in \cite{uhlmann,kalle},
is that the EoF is the \textit{convex closure} (or convex roof, as it is called in \cite{uhlmann})
of the pure state entanglement function
$E(\Psi) = S(\trace_A \Psi)$, restricted to the set of pure states.
This means that the \textit{epigraph} of the EoF
(being the set of points $(\rho,x)$ in $\cS(\cH)\times\R$ with $x\ge E_F(\rho)$)
on the complete state space $\cS(\cH)$ is
the convex closure of the epigraph of the function $E'$ defined over $\cS(\cH)$, where
$$
E'(\rho) = \left\{
\begin{array}{ll}
E(\rho),& \rho \mbox{ pure} \\
+\infty,& \rho \mbox{ not pure}. 
\end{array}
\right.
$$
This follows immediately from Cor. 17.1.5 of \cite{rock} and the definition (\ref{eq:def_eof}).
Note now that $E$ is concave over its domain. There is, therefore,
no need to explicitly exclude mixed states
\footnote{Of course, $E(\rho)$ has no real physical significance for mixed states. Moreover, we must
be careful to distinguish between the two possible definitions $E(\rho)=S(\trace_A\rho)$
and $E'(\rho)=S(\trace_B\rho)$. On pure states, these two definitions yield the same value, but for
mixed states this is not so anymore.},
so $E_F$ is the convex closure of $E$ as well.

In the following paragraphs we will apply the standard convex analytical formalism
for convex closures to general bounded functions $f$ whose domain is the convex set of states $\cS(\cH)$.
We will denote the convex closure of $f$ by $\hat{f}$.
One definition of the convex closure of $f$ is
\be \label{eq:def_conv}
\hat{f}(\rho) = \min_{\{(p_i,\rho_i)\}} \{ \sum_i p_i f(\rho_i) : \sum_i p_i \rho_i = \rho \},
\ee
agreeing, indeed, with the definition of the EoF.
A less cumbersome formulation of the convex closure is based on
Cor. 12.1.1 of \cite{rock}, which states that the convex closure of a function $f$
is the pointwise supremum of the collection of all affine functions on $\cS(\cH)$ majorised by $f$.
So, for all states $\rho$:
\be \label{eq:eof_dual}
\hat{f}(\rho) = \sup_{X\in \cB^s(\cH)} \{\trace\rho X: (\forall\psi\in\cH: \trace\Psi X \le f(\Psi))\}.
\ee
The mentioned affine functions are here the functions $\trace\Psi X$, where
$X$ ranges over $\cB^s(\cH)$
\footnote{For our purposes the Corollaries from \cite{rock} have to be restated with $\R^n$
replaced by $\cS(\cH)$. This causes no problems if one extends the domain of $f$ to
the affine space of all trace 1 \textit{Hermitian} operators and defines $f(x)=+\infty$ for negative $x$.
}.
This dual formulation is then further simplified by defining an intermediate function $f^*$:
\be
f^*(X) = \max_{\rho\in\cS(\cH)} \trace[\rho X]-f(\rho) \label{eq:g2},
\ee
the so-called \textit{conjugate function} of $f$ \cite{rock}.
If $f$ is continuous, then the conjugate function is just the Legendre transform of $f$.
The conjugate function is convex in $X$, because it is a pointwise maximum of functions that are affine
in $X$.
The importance of the conjugate function is that
the conjugate of the conjugate of $f$ is the convex closure of $f$, $\hat{f}=f^{**}$, and
the conjugate of the convex closure of $f$ is the conjugate of $f$, $\hat{f}^* = f^*$ (\cite{rock}, the remark
just before its Theorem 12.2).
Thus
\bea
\hat{f}(\rho) &=& \max_{X   \in \cB^s(\cH)} \trace[\rho X]-f^*(X)         \label{eq:eof_dual3} \\
f^*(X)        &=& \max_{\rho\in \cS(\cH)} \trace[\rho X]-\hat{f}(\rho). \label{eq:g3}
\eea
In other words, the conjugate and convex closure determine each other completely.

Because $f^*$ and $\hat{f}$ are convex functions, the optimal $X$ and $\rho$ in (\ref{eq:eof_dual3})
and (\ref{eq:g3}), respectively, both form convex sets (possibly singleton sets).
Furthermore, there is a correspondence between the optimal $X$ in (\ref{eq:eof_dual3})
and the optimal $\rho$ in (\ref{eq:g3}).
\begin{proposition}
(a) If $X'$ is an optimal $X$ for $\tau$ in (\ref{eq:eof_dual3}),
then (i) $\tau$ is an optimal $\rho$ for $X'$ in (\ref{eq:g3}),
and (ii) all members of an optimal
realising ensemble for $\tau$ are optimal $\rho$ for $X'$ in (\ref{eq:g2}).
(b) If $\rho'$ is an optimal $\rho$ for $Y$ in (\ref{eq:g3}),
then $Y$ is an optimal $X$ for $\rho'$ in (\ref{eq:eof_dual3}).
\end{proposition}
\textit{Proof.}
Statement (a)(i) is proven by inserting (\ref{eq:g3}) in (\ref{eq:eof_dual3}) and exploiting the
premise that $X'$ is an optimal $X$. This gives
$\hat{f}(\tau) = \trace\tau X' - \max_\rho(\trace\rho X'-\hat{f}(\rho))$.
Putting $\rho=\tau$ yields an upper bound on the right-hand side because $\tau$ is not necessarily
optimal in the maximisation.
However, the value of the bound we obtain is $\hat{f}(\tau)$, which happens to be equal to the left-hand side.
Thus this choice really is an optimal one, proving optimality of $\tau$ for $X'$ in (\ref{eq:g3}).

Statement (b) is proven similarly, by inserting (\ref{eq:eof_dual3}) in (\ref{eq:g3}).

Considering statement (a)(ii),
let $\{(p_i,\tau_i)\}$ be an optimal ensemble for $\tau$ (with $p_i>0$).
Thus $\hat{f}(\tau) = \sum_i p_i f(\tau_i)$.
By assumption, $\hat{f}(\tau) = \trace\tau X'-f^*(X')$. Inserting (\ref{eq:g2}) and expanding
unity as $\sum_i p_i$ yields
$\sum_i p_i f(\tau_i) = \trace\tau X' - \sum_i p_i \max_\rho(\trace\rho X'-f(\rho))$.
If we now replace $\rho$ by $\tau_i$ in the $i$-th summation term we get an upper bound on the
right-hand side, with equality only if all the $\tau_i$ are optimal $\rho$ for $X'$.
The bound is easily seen to be $\sum_i p_i f(\tau_i)$, which is actually equal to the left-hand side.
We find again that the bound is sharp, and optimality of the $\tau_i$ follows.
\qed
\section{Additivity} \label{sec:add}
These basic results will now prove to be a powerful tool for studying the additivity issue of the EoF.
Let $\cH_I$ and $\cH_{II}$ be two copies of the Hilbert space $\cH_A\otimes \cH_B$,
and define $\cH=\cH_I\ktimes\cH_{II}$.
We will reserve the symbol $\otimes$ for tensor products with respect to the A-B
subdivision, and the symbol $\ktimes$ for tensor products regarding
the I-II subdivision.
Strong superadditivity of the EoF \cite{kalle} is the inequality
\be \label{eq:ssa}
E_F(\rho) \ge?\, E_F(\rho_I) + E_F(\rho_{II}),
\ee
for $\rho$ a state on $\cH$, and $\rho_I$ and $\rho_{II}$ its reductions to $\cH_I$ and $\cH_{II}$, respectively.

The following Lemma is simple but crucial:
\begin{lemma}
For any bounded function $f$ defined on $\cS(\cH)$, strong superadditivity of $\hat{f}$
\be \label{eq:tbp}
\hat{f}(\rho) \ge?\, \hat{f}(\rho_I) + \hat{f}(\rho_{II}),
\ee
is equivalent to subadditivity of the conjugate function $f^*$ with respect to the Kronecker sum:
\be \label{eq:tbp2}
f^*(X_1\ktimes\identity+\identity\ktimes X_2) \le?\, f^*(X_1)+f^*(X_2).
\ee
\end{lemma}
\textit{Proof.}
Set $Z=X_1\ktimes\identity+\identity\ktimes X_2$.
Then, using (\ref{eq:eof_dual3})
and assuming the validity of (\ref{eq:tbp2}) yields
\beas
\hat{f}(\rho) &=& \sup_X \trace[\rho X]-f^*(X) \\
&\ge& \sup_{X_1,X_2}\trace[\rho Z]-f^*(Z) \\
&\ge& \sup_{X_1,X_2}\trace[\rho_I X_1+\rho_{II} X_2]-f^*(X_1)-f^*(X_2) \\
&=& \hat{f}(\rho_I)+\hat{f}(\rho_{II}),
\eeas
which is (\ref{eq:tbp}).

The converse follows from (\ref{eq:g3}).
Assuming the validity of (\ref{eq:tbp}) yields
\beas
f^*(Z) &=& \max_\rho \trace[\rho Z]-\hat{f}(\rho) \\
&\le& \max_\rho \trace[\rho_I X_1+\rho_{II} X_2]-\hat{f}(\rho_I)-\hat{f}(\rho_{II}) \\
&=& \max_{\rho_1,\rho_2} \trace[\rho_1 X_1+\rho_2 X_2]-\hat{f}(\rho_1)-\hat{f}(\rho_2) \\
&=& f^*(X)+f^*(Y),
\eeas
which is (\ref{eq:tbp2}).
\qed

The appearance of the Kronecker sum in Lemma 1 suggests that
the consideration of the function $f^*\circ\log$ is a more natural setting for
studying additivity.
Defining $g:=f^*\circ\log$ and setting $X_i=\log M_i$, (\ref{eq:tbp2}) becomes
$$
g(M_1\ktimes M_2) \le?\, g(M_1)+g(M_2),
$$
for $M_1,M_2\in\cB^+(\cH)$.
Restating (\ref{eq:eof_dual3}) and (\ref{eq:g3}) in terms of $M$, we have
\bea
g(M) &=& \max_{\rho\in\cS(\cH)} \trace[\rho \log(M)]-f(\rho) \label{eq:gs2} \\
\hat{f}(\rho) &=& \max_{M\in\cB^+(\cH)} \trace[\rho\log(M)] - g(M). \label{eq:fs}
\eea
Strictly speaking, these quantities are defined only for positive $M$.
However, when $M$ is singular, we can still make sense out of it by the usual extension $\trace[\rho\log(M)]=-\infty$
for any $\rho$ that is not completely supported on the range of $M$.

We can now restate Lemma 1 in the form of a Theorem, which is our first main result:
\begin{theorem}
For any function $f$ defined on $\cS(\cH)$,
and with $g$ defined on $\cB^+(\cH)$ by (\ref{eq:gs2}),
strong superadditivity of the convex closure $\hat{f}$,
\be \label{eq:tbpa}
\hat{f}(\rho) \ge?\, \hat{f}(\rho_I) + \hat{f}(\rho_{II}),
\ee
is equivalent to subadditivity of $g$,
\be \label{eq:tbp2a}
g(M_1\ktimes M_2) \le?\, g(M_1)+g(M_2).
\ee
\end{theorem}
Note that the expression $\trace[\rho\log(M)] - g(M)$ is invariant
under multiplication of $M$ by a positive scalar. Hence, one could impose the restriction $\trace M=1$,
i.e.\ that $M$ should be a \textit{state}, or alternatively $M\le\identity$, which is what we shall do.

An immediate corollary of this Theorem is the equivalence of the strong superadditivity of the EoF
with the subadditivity of $g=E^*\circ\log$, where $E^*$ is the conjugate of the entanglement functional
$E(\rho)=S(\trace_A\rho)$. We have chosen to present Theorem 1 in the more general way because
it obviates the rather remarkable independence of the Theorem on any property of the function $f$ at all.
Specifically, while for the sake of defining the EoF it is necessary to split up the Hilbert space into
two parties A and B, this is something the Theorem is completely oblivious of.

The only interesting feature of $E$ we can exploit at this level is its concavity. Concavity allows to
simplify the conjugation expression by replacing the maximisation over all mixed states by a maximisation
over pure states. Indeed, the argument of the maximisation
in
$$
g(M) = \max_{\rho\in\cS(\cH)} \trace[\rho \log(M)]-E(\rho)
$$
is a convex function of $\rho$, and it is well-known \cite{rock} that a convex function achieves its maximum
over a convex set always in an extreme point of that set, in this case in a pure state.
Thus:
$$
g(M) = \max_{\psi\in\cH} \trace[\Psi \log(M)]-E(\Psi).
$$

Theorem 1 reduces the additivity problem for the convex closure,
originally defined as a minimisation over \textit{ensembles},
to an equivalent problem for the conjugate function, defined as a maximisation over
\textit{pure states}.
If counterexamples are found for (\ref{eq:tbp2a}), this
automatically disproves strong superadditivity (\ref{eq:tbpa}), so this simplification does not
come at the cost of reduced power.
Specifically, by ``inverting'' the proof of Lemma 1 (or Theorem 1) and employing Proposition 1,
we easily get the following:
\begin{proposition}
If $\rho$ violates strong superadditivity of $\hat{f}$, (\ref{eq:tbpa}),
$M_1$ is optimal for $\rho_I$ in (\ref{eq:fs}),
and $M_2$ is optimal for $\rho_{II}$,
then $M_1\ktimes M_2$ violates subadditivity of $g$ (\ref{eq:tbp2a}).
If $M_1\ktimes M_2$ violates (\ref{eq:tbp2a}) and $\rho$ is optimal for $M_1\ktimes M_2$ in
(\ref{eq:gs2}), then $\rho$ violates (\ref{eq:tbpa}).
\end{proposition}
\section{Maximal Output Purity} \label{sec:mop}
Exploiting Theorem 1, we will now show that strong superadditivity of $E_F$ would follow as a
consequence of another additivity conjecture, concerning quantum channel capacities.
Recollect that, since $E$ is concave, the optimal $\rho$ in (\ref{eq:g2}) will be an extreme point of the
feasible set, i.e.\ a pure state, so:
\be
E^*(X) = \max_{\psi\in\cH} \trace[\Psi X]-E(\Psi). \label{eq:g}
\ee
From the additivity of $E$ over pure states it easily follows that the corresponding function
$g=E^*\circ\log$ is superadditive, hence subadditivity
of $g$ implies its additivity.

\subsection{Step 1}
The maximisation in $g$ can be rewritten in terms of a maximal eigenvalue $\lambda_{\max}$:
\begin{lemma}
For any $M\in\cB^+(\cH)$,
\bea
g(M) &:=& \max_\psi (\trace[\Psi \log M]-S(\trace_A\Psi)) \nonumber \\
&=& \max_{\tau\in\cS(\cH_B)} \lambda_{\max}(\log M+\log(\identity_A\otimes\tau)). \label{eq:g4}
\eea
\end{lemma}
Note that we will henceforth consider $\log M+\log(\identity_A\otimes\tau)$ as an operator restricted
to the range intersection $\ran(M)\cap\ran(\identity\otimes\tau)$.

\textit{Proof.}
\begin{eqnarray}
&&  \max_\tau \lambda_{\max}(\log M + \log \identity_A \otimes \tau ) \nonumber \\
&=& \max_\tau \max_\psi \trace[ \Psi (\log M + \log \identity_A \otimes \tau )] \label{RRstep} \\
&=& \max_\tau \max_\psi \trace[ \Psi \log M] + \trace[ \trace_A (\Psi) \log \tau ] \nonumber \\
&=& \max_\psi \trace[ \Psi \log M] -S(\trace_A \Psi) \label{relativeEntStep}.
\end{eqnarray}
In step (\ref{RRstep}) we have used the
Rayleigh-Ritz representation of a maximal eigenvalue, and in step
(\ref{relativeEntStep}) we have used the fact that relative
entropy is non-negative and attains the value zero when (and only when) its arguments
are equal. Specifically:
\beas
0 &=& \min_\tau S(\rho||\tau) \\
&=& \min_\tau -S(\rho)-\trace[\rho\log\tau] \\
&=& -S(\rho) -\max_\tau \trace[\rho\log\tau].
\eeas
\qed

\subsection{Step 2}
Using the Lie-Trotter formula, the logarithm can be replaced by a limit of a power function.
\begin{lemma}
$$
\exp g(M) = \lim_{p\rightarrow 0} h_p^{1/p}(M),
$$
where
$$
h_{p}(M) := \max_{\tau} ||M^{p/2} (\identity\otimes\tau)^p M^{p/2}||
$$
and $||.||$ denotes the operator norm.
\end{lemma}
\textit{Proof.}
Taking the exponential of both sides of (\ref{eq:g4}) and noting
$\exp\lambda_{\max}(M) = \lambda_{\max}\exp(M)$,
we get
$$
\exp g(M) = \max_\tau ||\exp(\log M+\log(\identity\otimes\tau))||.
$$
To make sense of this formula, we extend $\exp(\log M+\log(\identity\otimes\tau))$
as 0 on the complement of $\ran(M)\cap\ran(\identity\otimes\tau)$, as in \cite{hiaipetz93}.
The Lie-Trotter formula has a continuous version (see the remark after Lemma 3.3 in \cite{hiaipetz93})
$$
\exp(A+B) = \lim_{p\rightarrow 0} \big(\exp(pA/2) \exp(pB) \exp(pA/2)\big)^{1/p}.
$$
In particular, this gives us
\begin{equation} \label{eq:LT}
\exp(\log M+\log(\identity\otimes\tau))
= \lim_{p\rightarrow 0} \big(M^{p/2} (\identity\otimes\tau)^p M^{p/2}\big)^{1/p}.
\end{equation}

Define the shorthand functions
\beas
f(\tau)   &:=& ||\exp(\log M+\log(\identity\otimes\tau))|| \\
f_p(\tau) &:=& ||(M^{p/2} (\identity\otimes\tau)^p M^{p/2})^{1/p}||
\eeas
over $\cS(\cH)$.
By (\ref{eq:LT}) and the triangle inequality for norms, $f_p$ converges pointwise to $f$.
The functions $f_p$ are clearly continuous for $p>0$.
By Lemma 4.1 of \cite{hiaipetz93}, $f$ is continuous too.
From \cite{andohiai} (p.~118) we have that $f_p$ decreases monotonously to $f$ as $p$ decreases to 0.
The set $\cS(\cH)$, over which $f$ and $f_p$ are defined, is compact.
Hence, all the prerequisites are fulfilled to apply Dini's theorem \cite{apostol}, and we get that
the convergence of $f_p$ to $f$ is \textit{uniform} over $\cS(\cH)$.

Finally, uniform convergence is equivalent with convergence in the sup-norm. By
the triangle inequality for norms, that in turn implies that the sup-norm of
$f_p$ converges to the sup-norm of $f$. Therefore,
$h_p^{1/p}(M) = \max_\tau f_p(\tau) = ||f_p||_{\cS}$ converges to $||f||_{\cS}=\max_\tau f(\tau)=\exp g(M)$.
\qed

Additivity of $g$ would thus follow as a consequence of multiplicativity of $h_p$,
$h_p(M_1\ktimes M_2) =?\, h_p(M_1) h_p(M_2)$, for $p\downarrow 0$. Following \cite{amosov00},
we say that a property holds for $p\downarrow a$ if it holds for an
arbitrarily small, but finite, interval $p\in(a,a+\epsilon]$, $\epsilon>0$.

\subsection{Step 3}
The quantity $h_p(M)$ is formally equal to
the \textit{maximal output purity} \cite{amosov00,werner02,king02}
of quantum filtering operations. Indeed,
\beas
h_p(M) &=& \max_{\tau,\phi} \trace[\Phi(M^{p/2} (\identity\otimes\tau)^p M^{p/2})] \\
&=& \max_{\tau,\phi} \trace[\tau^p \trace_A[M^{p/2}\Phi M^{p/2}]] \\
&=& \max_\phi ||\trace_A[M^{p/2}\Phi M^{p/2}]||_q \\
&=& \nu_q(\Lambda),
\eeas
where $q=1/(1-p)$ and $||.||_q$ denotes the Schatten $q$-norm \cite{HJII},
and $\nu_q(\Lambda)$ is the maximal output purity measured by the Schatten $q$-norm
of the (non-trace preserving) operation
\be \label{eq:ch}
\Lambda: \rho\mapsto\Lambda(\rho) = \trace_A[M^{p/2}\rho M^{p/2}].
\ee
If this operation would be trace preserving, we would call it a channel.

\subsection{Step 4}
We now claim that there is no advantage in restricting attention to
operations of the form (\ref{eq:ch}).
It is of course true that the class of operations (\ref{eq:ch}) is rather specific.
They admit a Kraus representation
such that the block column matrix $(A_i)_i$ obtained by stacking the Kraus element matrices $A_i$ vertically,
equals $M^{p/2}$, which is a positive matrix.
Necessary conditions are that $\sum_i A_i^\dagger A_i=M^p$ (which is $\le\identity$) and
the input dimension of the operation should equal the output dimension times the number of elements.

However, as regards the maximal output purity question, these structural peculiarities
offer no additional mileage.
To see this, consider the specific case that $M$ is a partial isometry
$M=U\Sigma U^\dagger$, where $\Sigma=\ket{1}\bra{1}\otimes\identity_B$
and $U$ is any unitary, then
\beas
\nu_q(\Lambda) &=& \max_{\phi\in\cH} ||\trace_A[U\Sigma^{p/2} U^\dagger\Phi U \Sigma^{p/2} U^\dagger]||_q \\
&=& \max_{\phi'\in\cH} ||\trace_A[U\Sigma^{p/2} \Phi' \Sigma^{p/2} U^\dagger]||_q \\
&=& \max_{\phi''\in\cH_B} ||\trace_A[U(\ket{1}\bra{1}\otimes\Phi'')U^\dagger]||_q,
\eeas
which is the generic case for operations from $\cH_A$ to $\cH_A$. Thus, the case for the ``special operations''
$\cH\mapsto\cH_A$ contains the generic $\cH_A\mapsto\cH_A$ case and is therefore not easier to prove.

\subsection{Step 5}
The exponent $p$ of $M$, occurring in $\Lambda$,
is coupled to $q$, occurring in $\nu_q$, via the relation $q=1/(1-p)$.
To cap off our argument, we ``decouple'' $p$ and $q$ by replacing $M^{p/2}$ with a
general matrix $0\le X\le\identity$, strenghtening our multiplicativity conjecture ever so slightly.
Noting finally that $p\downarrow 0$ corresponds to $q\downarrow 1$,
we get our second main result:
\begin{theorem}
If $\nu_q(\Lambda)$ is multiplicative for $q\downarrow 1$
and for any filtering operation $\Lambda$,
then the entanglement of formation is strongly subadditive.
\end{theorem}

Multiplicativity of $\nu_q$ had been conjectured in \cite{amosov00} for trace preserving channels.
It has been proven for entanglement breaking channels \cite{king02},
unital qubit maps \cite{king3} and depolarising channels \cite{king4},
but, unfortunately, was refuted in \cite{werner02} for $q>4.79$.
Nevertheless, the conjecture might still be true for $q\downarrow 1$.

Theorem 2 has to be compared to the main technical result in \cite{winter02}, which states that additivity
of the Holevo capacity for given channels implies additivity of the EoF for certain states.
In a sense, our Theorem 2 is stronger because we get the stronger outcome of strong subadditivity.
On the other hand, this comes at the price of having to consider non-trace-preserving operations.

After the appearance of the first draft of this manuscript, Shor proved \cite{shor}
the equivalence of four additivity conjectures: strong superadditivity of the EoF,
ordinary additivity of the EoF, additivity of the maximal output purity $\nu_S$ of a channel as measured
by the entropy, and additivity of the classical (Holevo) capacity of a channel.
As multiplicativity of $\nu_q(\Lambda)$ for $q\downarrow 1$ implies additivity of
$\nu_S(\Lambda)$ \cite{amosov00}, Shor's third equivalence provides an alternative proof
for our result Theorem 2.
\section{Conclusion} \label{conclusion}
In conclusion, we have shown how a simple convex analytical argument leads to a simpler formulation of the entanglement
of formation and an especially simple equivalent condition for strong superadditivity of the EoF.
Based on this we have found the second result that strong superadditivity of the EoF
would follow as a consequence of the multiplicativity of the maximum output purity $\nu_q$ of
quantum filtering operations, for $q\downarrow 1$.
\begin{acknowledgement}
We gratefully acknowledge comments by M.B. Plenio, J. Eisert, M.B. Ruskai and Ch. King.
SLB currently holds a Wolfson-Royal Society Research Merit Award.
\end{acknowledgement}

\end{document}